# Research on Pd film deposition rate calculation and simulation based on TiZrV/Pd film coating experiment *

Jie Wang, Bo Zhang#, Yanhui Xu, Yong Wang

National Synchrotron Radiation Laboratory，University of Science and Technology of China, HeFei, AnHui 230029 China

* Financially supported by the National Natural Science Funds of China (Grant No. 11205155) and Fundamental Research Funds for the Central Universities (WK2310000041).

# corresponding author: zhbo@ustc.edu.cn；phone numbers: +8613615691450

**ABSTRACT**

The vacuum chamber of accelerator storage ring need clean ultra-high vacuum environment. TiZrV getter film which was deposited on interior wall of vacuum chamber, can realize distributed pumping, effectively improve the vacuum degree and reduce the longitudinal gradient. But accumulation of pollutants such as $N_2$, $O_2$, will decrease the adsorption ability of non-evaporable getter (NEG), which leads to the reduction of NEG lifetime. Therefore, NEG thin film coated with a layer of Pd which has high diffusion rate and absorption ability for $H_2$, can extend the service life of NEG, and improve the pumping rate of $H_2$ at the same time. With argon as discharge gas, magnetron sputtering method was adopted to prepare TiZrV-Pd film in long straight pipe. According to the experimental results of the scanning electron microscope (SEM), deposition rates of TiZrV-Pd films were analyzed under different deposition parameters, the magnetic field strength, the gas flow rate, discharge current, discharge voltage and working pressure. Moreover, comparing the simulation results based on Sigmund's theory and experimental results, it was shown that the deposition rate *C* can be estimated by the depth sputtered, *D* for Pd film coatings in this experiment device.

**INTRODUCTION**

In order to reduce beam losses caused by residual gas scattering, the storage ring needs clean ultra-high vacuum environment, to maintain a long beam lifetime. The ways to obtain and maintain ultra-high vacuum are mainly through discrete distribution of vacuum pumps, such as sputtering ion pumps and titanium sublimation pumps. However, the synchrotron radiation and high energy particles bombarding the inner wall surface will cause surface outgassing and high dynamic gas load. At the same time, because of the limitation of the vacuum chamber conductance, vacuum degree near vacuum pumps is better than that far away from the vacuum pumps. TiZrV non-evaporable getter (NEG) film which was deposited on the inner wall of vacuum chamber, can realize distributed pumping effectively, improve the vacuum degree and reduce the longitudinal gradient. Nowadays, TiZrV film has been applied in accelerator field. TiZrV has low secondary electron yield, and can adsorb many gases, such as $O_2$, $N_2$, CO, $CO_2$. Nevertheless, with repeated exposure to air, TiZrV film will be polluted by $N_2$, $O_2$ and so on, which can reduce the absorbing behavior and shorten the life of TiZrV film.

When the pressure is less than $1\times10^{-9}$ mbar, the main components of the residual gas are $H_2$. In order to obtain ultra-high vacuum (UHV), vacuum pump with a high pumping speed to $H_2$ is needed. Therefore, C. Benvenuti et al. [1] put forward that Pd film with a high adsorption of $H_2$, can be added on NEG film. At the same time, Pd film can protect NEG film and prolong the service life of NEG thin films. In addition, the research results by M. Mura et al. [2] showed that TiZrV-Pd film can improve the adsorption factor for $H_2$ and other gases with high adsorption factor. So they applied TiZrV-Pd films to ion pumps and obtained better vacuum performance. In addition, the research results by C. Benvenuti et al. showed that TiZrV-Pd film can prolong the service life of NEG film, also improve the pumping speed for $H_2$ and CO, but not for $N_2$ and $CO_2$. The secondary electron yield of TiZrV-Pd film caused by synchrotron radiation is lower than TiZrV films.

However, in the process of film coating, the effect of magnetic field strength, gas flow rate, discharge current, discharge voltage, working pressure on the deposition rates of TiZrV-Pd film are rarely mentioned. So, we studied the effects of these parameters on the deposition rate of TiZrV-Pd film, by magnetron sputtering deposition method.

## APPARATUS AND METHODS

This experiment adopts magnetron sputtering coating method. Coated pipe is Φ 86 mm×540 mm stainless steel pipe. As shown in Fig. 1, the coating system mainly include pipe to be coated, molecular pump system, gas control system, discharge power, cathode wires, sample holder and solenoid. The cathode targets had two types: one is twisted wires by Ti, Zr, V wires with diameter of 2.5 mm, the other is a 1 mm diameter Pd wire. This deposition system has been detailed introduced in reference [3]. Film thickness was measured by use of a Sirion 200 Schottky field SEM.

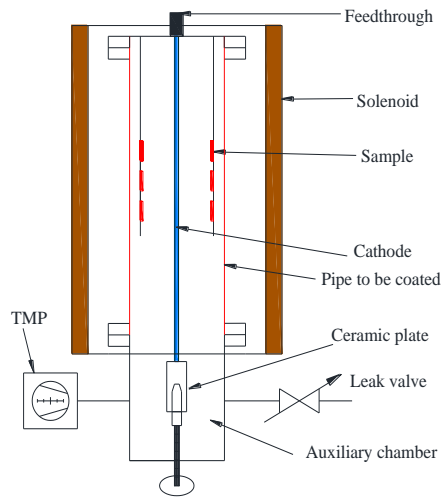

Figure 1: Schematic diagram of sputtering coating system.

**Theoretical calculation**

The deposition rate calculation and transport of sputtered particles from the target towards the film are problems long discussed in the literatures in the past several decades [4, 5, 6]. Some of the models which have been developed to simulate the evolution of surfaces in three dimensions are based on continuum equations, which have provided useful information on the morphology of step coverage for various amounts of surface diffusion and angular distributions [7]. Other models which mainly rely on a kinetic Monte Carlo model, the "Effective Thermalizing Collision" (ETC) approximation has successfully explained the so-called Keller-Simmons (K-S) formula, using empirical equation in classical magnetron sputtering deposition with a single sputter gas [8]. However, these models are time-consuming, less portable and complicated. According to the comparison between experiment and simulation results, we found that the deposition rate of single component Pd metal film can be estimated by sputtering depth for cylindrical pipe coating.

The depth sputtered, $D$, and the sputtering yield, $Y$ are directly related via [9]:

$$D = \frac{JYr^3t}{e_0}$$

where $J$ is the beam current density, $Y$ is the sputtering yield, $t$ is the time of sputtering, $r^3$ is the volume of an atom of the sample and $e_0$ the charge on the electron. Certainties in $Y$ lead directly to certainties in the thickness scale, $D$. There have been a number of predictive relations to enable $Y$ to be calculated, based on Sigmund's theory [10]. In this article, two approaches used are those of Matsunami et al. [11] and the later development by Yamamura and Tawara [12]. Deposition rate $C$ by experiment: $C = T/t$, where $T$ is the thickness of film, $t$ is the deposition time. According to reference [13-15], the error of the experimental measuring values are about 5%. Analysis and comparison have been done between the experimental data for argon ion deposition yield of Pd film and semi-empirical formulas based on Matsunami et al.'s and Yamamura et al.'s theories in the energy range 350-700 eV.

Adopting Wolfram Mathematica software, deposition rate of Pd film for the cylindrical pipe is calculated by two models, Yamamura et al. and Matsunami et al., in most cases, the margin of error between experimental value and the simulation values was -7% ~ 6% for Yamamura et al.'s model, and 2% ~ 11% for Matsunami et al.'s model, as shown in Fig. 2. But in individual cases, the margin of error can reach 16% for both models. By comparing the calculation results, it was found that the simulation results of Yamamura and Tawaras model is closer to the experimental results.

Amorphization, damage and implanted ions all occur in the early stage of sputtering and their effects were included in the revised parameter $Q$ in M. P. Seah et al. model [9]. Based on M. P. Seah et al. -- Matsunami et al. model, the margin of error between experimental data and simulation values was -2.5% ~ 20%, shown in Fig. 3. For M. P. Seah et al. -- Yamamura and Tawaras model, it was -7.7% ~ 10%. Particularly, for sample #1, #5, #6, #7, the margin of error are 0 %, -1.6 %, 0.8 %, 1.6 % respectively. However, for sample #2, #3, #4, they are -7.7 %, -7.7 %, 10 %.

Comparing the simulation results and experimental results, it was shown that the deposition rate $C$ and the depth sputtered, $D$, are the same in most cases. So the deposition rate $C$ can be estimated by the depth sputtered, $D$ for Pd film in this experiment equipment.

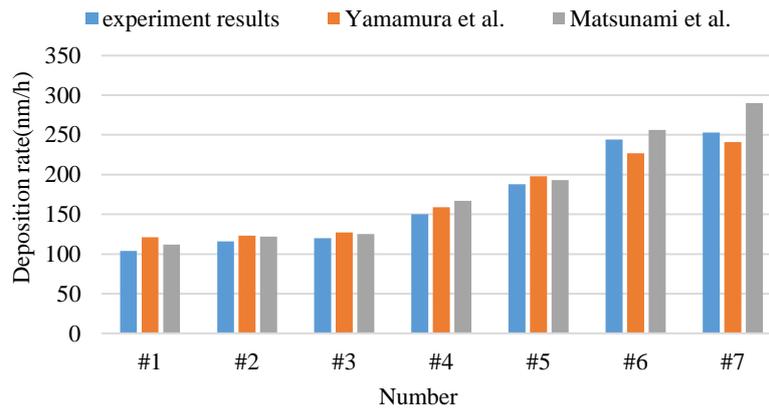

Figure 2: Comparison between experiment results and calculation results of two models, Yamamura et al. and Matsunami et al.

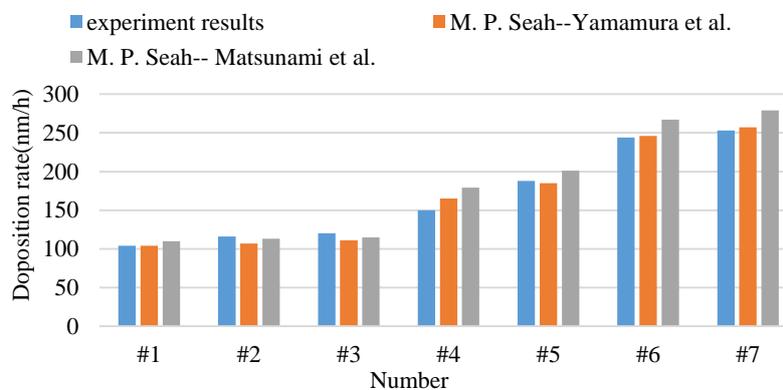

Figure 3: Comparison between experiment results and calculation results of M. P. Seah et al. model.

**Experimental results and discussion**

According to the test result of SEM, the effects of work pressure, gas flow, magnetic field strength, the discharge current on deposition rates of TiZrV-Pd films were analyzed.

*The influence of working pressure on deposition rates*

For TiZrV film on silicon substrate, the discharge current of 0.2 A, gas flow rate of 2.0 Sccm, magnetic field strength of 120 Gauss remains unchanged, under the condition of working pressure changing between 1 ~ 20 Pa. When work pressure was 2 Pa, the deposition rate of TiZrV was the highest, about 266 nm/h, and the corresponding discharge voltage is the highest, shown in table 1.

Table 1：The effect of working pressure on the deposition rate of TiZrV, with 0.2A discharge current, 2.0 Sccm gas flow rate, 120 Gauss magnetic field strength.

| Sample | working pressure /Pa | deposition rate /nm/h | discharge voltage /V |
|---|---|---|---|
| #0610 | 1 | 260 | 394-365 |

| Sample | | | |
|---|---|---|---|
| #0311 | 2 | 266 | 427-436 |
| #0527 | 5 | 147 | 334-319 |
| #0604 | 10 | 170 | 350-314 |
| #0612 | 20 | 139 | 350-329 |

For Pd film with TiZrV film substrate, 0.04 A discharge current, 2.0 Sccm gas flow, 175 Gauss magnetic field strength remained unchanged, working pressure changing between 2 ~ 20 Pa. When working pressure was 2 Pa, the deposition rate of Pd film was highest, about 240 nm/h, the corresponding discharge voltage is the highest, shown in table 2.

Table 2：The effect of working pressure on the deposition rate of Pd, with 0.02A discharge current, 2.0 Sccm gas flow rate, 120 Gauss magnetic field strength.

| Sample | working pressure /Pa | deposition rate /nm/h | discharge voltage /V |
|---|---|---|---|
| #0227 | 2 | 244 | 530-523 |
| #0618 | 5 | 166 | 437-548 |
| #0508 | 10 | 186 | 456-527 |
| #0620 | 20 | 166 | 464-573 |

*The influence of gas flow on deposition rates*

For TiZrV film with silicon substrate, 0.25 A discharge current, 2.0 Pa working pressure, 123 Gauss magnetic field strength remained unchanged, gas flow changing between 1 ~ 4 Sccm, shown in table 3. The highest deposition rate of TiZrV film was 278 nm/h, when the gas flow was 1 Sccm. The influence of gas flow on the deposition rate connected with the discharge voltage. Hence, a higher discharge voltage contribute to the improvement of the deposition rate of TiZrV film. It was found the same results for Pd film shown in table 4. When the gas flow rate was 5 Sccm, the highest deposition rate of Pd film was about 240 nm/h.

Table 3：The effect of gas flow on the deposition rate of TiZrV, with 0.25A discharge current, 2.0 working pressure, 123 Gauss magnetic field strength.

| Sample | gas flow /Sccm | deposition rate /nm/h | discharge voltage /V |
|---|---|---|---|
| #0617 | 1 | 278 | 350-400 |
| #0619 | 3 | 233 | 351-375 |
| #0624 | 4 | 270 | 371-385 |

Table 4：The effect of gas flow on the deposition rate of Pd, with 0.02A discharge current, 2.0 working pressure, 175 Gauss magnetic field strength.

| Sample | gas flow /Sccm | deposition rate /nm/h | discharge voltage /V |
|---|---|---|---|
| #0606 | 1 | 177 | 424-458 |
| #1225 | 2 | 240 | 664-697 |
| #0611 | 3 | 160 | 440-446 |
| #0613 | 4 | 146 | 436-465 |
| #0225 | 5 | 210 | 571-574 |

*The influence of magnetic field strength on deposition rates*

For TiZrV film with silicon substrate, 2.0 Pa working pressure, 2.0 Sccm gas flow, 0.2 A discharge current remained unchanged, under the condition of magnetic field strength changing between 80 and 200 Gauss, shown in table 5. When the magnetic field strength was 82 Gauss, the highest deposition rate of TiZrV film was about 500 nm/h. The influence of gas flow on the deposition rate connected with the magnetic field strength. Hence, a higher discharge voltage contribute to improve the deposition rate of TiZrV film. It was found the same results for Pd film shown in table 6. When the magnetic field strength was 175 Gauss, the highest deposition rate of Pd film was about 240 nm/h.

Table 5: The effect of magnetic field strength on the deposition rate of TiZrV film, with 0.2A discharge current, 2.0 working pressure, 2.0 Sccm gas flow.

| Sample | magnetic field strength /Guass | deposition rate /nm/h | discharge voltage /V |
|---|---|---|---|
| #0409 | 82 | 500 | 606-658 |
| #0402 | 93 | 490 | 508-561 |
| #0331 | 107 | 317 | 394-471 |
| #0311 | 123 | 266 | 427-436 |
| #0521 | 177 | 240 | 381-308 |

Table 6: The effect of magnetic field strength on the deposition rate of Pd film, with 0.02A discharge current, 2.0 working pressure, 2.0 Sccm gas flow.

| Sample | magnetic field strength /Guass | deposition rate /nm/h | discharge voltage /V |
|---|---|---|---|
| #0514 | 65 | 163 | 443-440 |
| #0304 | 123 | 212 | 559-569 |
| #1225 | 175 | 240 | 664-697 |
| #0418 | 206 | 130 | 430-440 |

*The influence of discharge current on deposition rates*

For TiZrV film with silicon substrate, 175 Gauss magnetic field strength, 2.0 Pa working pressure, 2.0 Sccm gas flow remained unchanged, under the condition of discharge current changing between 0.1 and 0.5 A, shown in table 7. When the discharge current was 0.5 A, TiZrV thin film deposition rate was the highest, about 316 nm/h. According to the magnetron sputtering model [16] proposed by Thornton and Penfold, the relationship between ions current $\overline{J_i}$ and discharge current $I_{dc}$ is shown in the following equation (1):

$$\overline{J_i} = \frac{I_{dc}}{2\pi R w} \tag{1}$$

where $R$ is the average radius of high density plasma torus with a bright glow close to the cathode, $w$ is the width of high density plasma torus. Ions current will increase with the discharge current. According to equation (2), film deposition rate $R_{sput}$ will increase with ion current. Table 7 shows positive relationship between the deposition rate and discharge current but it is not linear. Because when the discharge current increases, $R$ and $w$ will change slightly. In short, theoretical analysis is consistent with the experimental results.

$$R_{sput} = \gamma_{sput} \frac{\overline{J_i}}{e} \frac{1}{n} \qquad (2)$$

where $\gamma_{sput} \sim 1$, $n$ is the atomic density of target material. It was found the same results for the influence of discharge current on Pd film deposition rates. When the discharge current was 0.03 A, the deposition rate of Pd film was highest, about 217 nm/h.

Table 7: The effect of discharge current on the deposition rate of TiZrV film, with 2.0 working pressure, 2.0 Sccm gas flow, 175 Gauss magnetic field strength.

| Sample | magnetic field strength /Guass | deposition rate /nm/h | discharge voltage /V |
|---|---|---|---|
| #1224 | 0.1 | 30 | 473-490 |
| #0108 | 0.25 | 80 | 547-557 |
| #0226 | 0.5 | 316 | 492-343 |

## CONCLUSION

The experiment results show that magnetic field strength, gas flow, discharge current and working pressure will affect the deposition rates of TiZrV-Pd film. Adjusting the values of above four factors and other experimental conditions, the maximum deposition rate of TiZrV thin film was 490 nm/h, it was about 240 nm/h for Pd film. The results meet the demand of actual coating requirement. Moreover, comparing the simulation results based on Sigmund's theory and experimental results, it was shown that the error range is within 10% for M. P. Seah et al. -- Yamamura and Tawaras model and the deposition rate $C$ can be estimated by the depth sputtered, $D$ for Pd film.